
\def\gl{\mathrel{\raise1ex\hbox{$>$\kern-.75em\lower1ex\hbox{$<$}}}}
\def\lg{\mathrel{\raise1ex\hbox{$<$\kern-.75em\lower1ex\hbox{$>$}}}}
\def\gtwid{\mathrel{\raise.3ex\hbox{$>$\kern-.75em\lower1ex\hbox{$\sim$}}}}
\def\ltwid{\mathrel{\raise.3ex\hbox{$<$\kern-.75em\lower1ex\hbox{$\sim$}}}}
\def\sqr#1#2{{\vcenter{\hrule height.#2pt
      \hbox{\vrule width.#2pt height#1pt \kern#1pt
         \vrule width.#2pt}
      \hrule height.#2pt}}}

\overfullrule=0pt

\def\eg{\hbox{{\it e.\ g.}}}\def\ie{\hbox{{\it i.\ e.}}}


\def\leaderfill{\leaders\hbox to 1em{\hss.\hss}\hfill}


\def\ref#1{${}^{#1}$}
\newcount\eqnum \eqnum=0  
\newcount\eqnA\eqnA=0\newcount\eqnB\eqnB=0\newcount\eqnC\eqnC=0\newcount\eqnD\eqnD=0
\def\eqnoi{\global\advance\eqnum by 1\eqno(\the\eqnum)}
\def\eqnai{\global\advance\eqnum by 1\eqno(\the\eqnum{a})}
\def\eqnbi{\eqno(\the\eqnum{b})}
\def\eqnoA{\global\advance\eqnA by 1\eqno(A\the\eqnA)}
\def\eqnoB{\global\advance\eqnB by 1\eqno(B\the\eqnB)}
\def\eqnoC{\global\advance\eqnC by 1\eqno(C\the\eqnC)}
\def\eqnoD{\global\advance\eqnD by 1\eqno(D\the\eqnD)}
\def\back#1{{\advance\eqnum by-#1 Eq.~(\the\eqnum)}}
\def\backs#1{{\advance\eqnum by-#1 Eqs.~(\the\eqnum)}}
\def\backn#1{{\advance\eqnum by-#1 (\the\eqnum)}}
\def\backA#1{{\advance\eqnA by-#1 Eq.~(A\the\eqnA)}}
\def\backB#1{{\advance\eqnB by-#1 Eq.~(B\the\eqnB)}}
\def\backC#1{{\advance\eqnC by-#1 Eq.~(C\the\eqnC)}}
\def\backD#1{{\advance\eqnD by-#1 Eq.~(D\the\eqnD)}}
\def\last{{Eq.~(\the\eqnum)}}                   
\def\lasts{{Eqs.~(\the\eqnum)}}                   
\def\lastn{{(\the\eqnum)}}                      
\def\lastA{{Eq.~(A\the\eqnA)}}\def\lastB{{Eq.~(B\the\eqnB)}}
\def\lastC{{Eq.~(C\the\eqnC)}}\def\lastD{{Eq.~(D\the\eqnD)}}
\newcount\refnum\refnum=0  
\def\refi{\smallskip\global\advance\refnum by 1\item{\the\refnum.}}

\newcount\rfignum\rfignum=0  
\def\rfigi{\medskip\global\advance\rfignum by 1\item{Figure \the\rfignum.}}

\newcount\fignum\fignum=0  
\def\figi{\global\advance\fignum by 1 Fig.~\the\fignum}

\newcount\rtabnum\rtabnum=0  
\def\rtabi{\medskip\global\advance\rtabnum by 1\item{Table \the\rtabnum.}}

\newcount\tabnum\tabnum=0  
\def\tabi{\global\advance\tabnum by 1 Table~\the\tabnum}

\newcount\secnum\secnum=0 
\def\chap#1{\global\advance\secnum by 1
\bigskip\centerline{\bf{\the\secnum}. #1}\smallskip\noindent}

\def\pd#1#2{{\partial #1\over\partial #2}}      
\def\p2d#1#2{{\partial^2 #1\over\partial #2^2}} 
\def\pnd#1#2#3{{\partial^{#3} #1\over\partial #2^{#3}}} 
\def\td#1#2{{d #1\over d #2}}      
\def\t2d#1#2{{d^2 #1\over d #2^2}} 


\def\ith{{$i^{\rm th}$}}

\def\2kth{{$2k^{\rm th}$}}

\def\n-th{{$(n-1)^{\rm th}$}}

\def\N-th{{$(N-1)^{\rm th}$}}

\def\0th{$0^{\rm th}$}
\def\1st{$1^{\rm st}$}
\def\2nd{$2^{\rm nd}$}
\def\3rd{$3^{\rm rd}$}
\def\4th{$4^{\rm th}$}
\def\5th{$5^{\rm th}$}
\def\5th{$6^{\rm th}$}
\def\6th{$7^{\rm th}$}
\def\7th{$7^{\rm th}$}
\def\8th{$8^{\rm th}$}
\def\9th{$9^{\rm th}$}

\def\a{{\alpha}}
\def\b{{\beta}}

\def\d{\delta}\def\D{\Delta}

\def\n{\nu}
\def\p{\pi}

\def\t{\tau}

\def\addr{
    \centerline{Center for Polymer Studies and Department of Physics}
     \centerline{Boston University, Boston, MA~ 02215}}

\def\cp #1 #2 #3 {{\sl Chem.\ Phys.} {\bf #1}, #2 (#3)}
\def\jcp #1 #2 #3 {{\sl J.\ Chem.\ Phys.} {\bf #1}, #2 (#3)}
\def\jfm #1 #2 #3 {{\sl J. Fluid Mech.} {\bf #1}, #2 (#3)}
\def\jpa #1 #2 #3 {{\sl J. Phys.\ A} {\bf #1}, #2 (#3)}
\def\jsp #1 #2 #3 {{\sl J. Stat.\ Phys.} {\bf #1}, #2 (#3)}
\def\pra #1 #2 #3 {{\sl Phys.\ Rev.\ A} {\bf #1}, #2 (#3)}
\def\prb #1 #2 #3 {{\sl Phys.\ Rev.\ B} {\bf #1}, #2 (#3)}
\def\pre #1 #2 #3 {{\sl Phys.\ Rev.\ E} {\bf #1}, #2 (#3)}
\def\prl #1 #2 #3 {{\sl Phys.\ Rev.\ Lett.} {\bf #1}, #2 (#3)}
\def\rmp #1 #2 #3 {{\sl Rev.\ Mod.\ Phys.} {\bf #1}, #2 (#3)}
\def\usp #1 #2 #3 {{\sl Sov.\ Phys. -- Usp.} {\bf #1}, #2 (#3)}

\overfullrule=0pt
\magnification 1200
\baselineskip=16 true bp

\def\dx{\Delta x}

\def\ab{$A+B\to 0$}
\def\td{$t^{-1/4}$}
\def\t13{$t^{-1/3}$}

\centerline{\bf KINETICS OF $A+B \to 0$ WITH DRIVEN DIFFUSIVE MOTION}
\vskip 0.2in
\centerline{I.~Ispolatov, P.~L.~Krapivsky, and S.~Redner}
\addr
\vskip 0.7in

\centerline{ABSTRACT}
{\narrower\bigskip\smallskip\noindent We study the kinetics of
two-species annihilation, \ab, when all particles undergo strictly
biased motion in the same direction and with an excluded volume
repulsion between same species particles.  It was recently shown that
the density in this system decays as \t13, compared to \td\ density
decay in \ab\ with isotropic diffusion and either with or without the
hard-core repulsion.  We suggest a relatively simple explanation for
this \t13 decay based on the Burgers equation.  Related properties
associated with the asymptotic distribution of reactants can also be
accounted for within this Burgers equation description.\bigskip}

\bigskip\bigskip
P. A. C. S. Numbers: 02.50.-r, 05.40.+j, 82.20.-w, 82.20.Wt

\vfil\eject

\chap{INTRODUCTION}

While the kinetics of two species annihilation, \ab, has been
extensively investigated and reasonably well understood [1-5],
surprising results have been recently obtained in one dimension when
both species move according to driven diffusive motion [6].  In this
model, each particle can attempt a move only to the right and the
attempt is successful only if the particle lands on an unoccupied
site.  If the particle lands on a site already occupied by a member of
the same species, the move attempt is rejected and the initial
particle remains at its starting position.  However, if the particle
lands on a site which is occupied by a member of the opposite species,
then an \ab\ reaction occurs and both particles disappear.  This
motion of each species may be viewed as strongly biased diffusion with
a hard core repulsion between same species particles.

For this system, one might naively anticipate that, by performing a
Galilean transformation to a zero momentum reference frame, the
long-time kinetics should be the same as that of \ab\ with isotropic
diffusion.  For this latter system, the density is known to decay in
time as $t^{-1/4}$ when the initial densities of the two species are
equal [1-5].  This result has been found to hold both for the case of
free diffusion and when an excluded-volume repulsion between same
species particles exists.  In fact such a Galilean universality has
been derived for the case of single species reactions and a special
type of two-species reaction [6].

However, extensive numerical simulations in one dimension indicate,
rather surprisingly, that for two-species annihilation with driven
motion, the density asymptotically decays as $t^{-1/3}$ for a Poissonian
initial distribution of the two species [7].  It was also found that
the approach of the density to the \t13 asymptotic form was rather
slow.  Even after $10^6$ time steps, local estimates of the density
decay exponent were still varying systematically with time.  On the
other hand, for isotropic diffusion, stable estimates of the decay
exponent are achieved after $10^3$ -- $10^4$ time steps.  As we
discuss below, there is an interesting reason for this slow approach
to asymptotic behavior in the driven system.

In this paper we provide a theoretical argument for the $t^{-1/3}$
decay of the density in \ab\ with driven diffusive motion.  Our
approach is based on first elucidating the dynamics of a single $AB$
interface when the two species are initially separated (Sec.\ II).  A
continuum description in terms of the inviscid Burgers equation,
together with continuity of particle currents, shows that the velocity
of the interface can be determined in terms of the relative densities
of the two adjoining species.  This information is used to infer that
when the initial condition is homogeneous and random, a typical
single-species domain grows as $t^{2/3}$.  This result, together with
the fact that the density inside a domain is simply related to initial
density fluctuations, leads to the $t^{-1/3}$ decay of the density.
In Sec.\ III, we present numerical simulations of the reaction,
focusing, in particular, on the density decay exponent, the domain
length growth exponent, and the behavior of the scaled domain density
profile.  This latter consideration has previously been found to be
fruitful in understanding both the kinetics and the spatial structure
of reactants in \ab\ with unbiased diffusion of each species [8].  Our
numerical results support many of our qualitative arguments.  Finally,
in Sec.\ IV, we provide general conclusions as well as discuss the
kinetics with arbitrary biases of each species and in higher spatial
dimension.

\chap{ANALYTICAL APPROACHES}
\centerline{\bf {A. Single species motion}}\smallskip

Let us first recall some basic properties of single-species driven
diffusion in one dimension where the positions of individual particles
are updated sequentially and in random order [9].  Denote by $n_i(t)$
the occupation number for the \ith\ site at time $t$.  By the
excluded-volume interaction, $n_i$ is restricted to be either 0 or 1.
Depending on the local configuration of particles on sites $i-1$, $i$,
and $i+1$, the occupation number for site $i$ at time $t+\d t$ has the
following three possible outcomes,
\smallskip
$$
n_i(t+\d t) =\cases{n_i(t) & with probability \quad $1-2\a\d t$,\cr
\cr
n_i(t) n_{i+1}(t) & with probability \quad ~$\a\d t$, \cr
\cr
n_i(t) + (1-n_i(t))n_{i-1}(t) & with probability \quad ~$\a\d t$. \cr}
\eqnoi
$$
\smallskip\noindent
Here $\d t$ is the microscopic time step and $\a\d t$ is the
probability that a particle attempts to move during the time step.
These three possibilities have the following origins: If neither the
\ith\ nor the $(i-1)^{\rm th}$ particles were selected for a move
attempt, then $n_i(t)$ does not evolve, as indicated in the first
line.  The second line describes the state of site $i$ at $t+\d t$ if
a move attempt is performed on the particle initially at site $i$.
Similarly, the third line gives the state of site $i$ if a move
attempt is performed on a particle initially at site $i-1$.

Corresponding to these elemental updates, the Master equation for the
average occupation number $\overline n_i(t)$ is,
$$
\overline n_i(t+\d t)- \overline n_i(t)=
\a\d t\,[\overline n_i(t)\overline n_{i+1}(t)
-\overline n_i(t)\overline n_{i-1}(t)
+\overline n_{i-1}(t)-\overline n_i(t)].
\eqnoi
$$
In writing this equation, we have exploited the fact that for
one-dimensional driven diffusion in a finite ring or in an open system
but sufficiently far from the boundaries, the occupation numbers for
the different sites are uncorrelated [9].  This implies that a
mean-field description is exact so that the two-particle correlation
function, $\overline{n_i(t)n_j(t)}$, may be replaced by the product of
single-particle densities, $\overline n_i(t)\overline n_j(t)$, in
\last.  In the continuum limit, the leading order expansion of the
above Master equation yields the Burgers equation,
$$
\pd {\overline n} t
= -{{ \dx} \over {\D t}} \pd{}{x} [\overline n (1-\overline n)] +
{{\dx ^2} \over {2 \D t}} \pnd{\overline n}{x}{2},
\eqnoi
$$
where $\dx$ is the lattice spacing and $\D t= \a^{-1}$ is a
macroscopic time step during which one move attempt is performed, on
average, for each particle. In the following we choose the length and
time scales such that $\dx=1$ and $\D t=1$.

This Burgers equation may be solved by first reducing it to the
diffusion equation by a Cole-Hopf transformation [10].  Since the
latter equation is trivially soluble, the solution to the Burgers
equation may be obtained by performing an inverse Cole-Hopf
transformation on the corresponding diffusive solution.  A sketch of
the Burgers equation solution in the long-time limit for a localized
initial source $\overline n(x,0) = A \d (x)$ is shown in Fig.~1.
There are three universal regions in the long-time solution which are
independent of the microscopic details of the initial density
distribution.  First, there is a constant-slope ``interior'' region
where $\overline n(x,t)\approx(1-x/t)/2$,with $x$ in the domain
$t-2\sqrt{At}\leq x\leq t$.  This behavior arises because the leading
low-density particles have a relatively large velocity and continue to
propagate faster than the denser trailing particles.  At either end of
this interior region, \ie, at $x=t-2\sqrt{A t}$ and at $x=t$, there
are two diffusive boundary regions whose widths grow as
$\sim\sqrt{t}$.  In the interior region, the solution to \last\ is
asymptotically close to the profile of the solution to the inviscid
Burgers equation (shown dotted in the figure).  This inviscid solution
consists of the straight line segment $\overline n_{\rm
inviscid}(x,t)=(1-x/t)/2$ for $t- 2\sqrt{A t} < x < t$, while
$\overline n_{\rm inviscid}(x,t)=0$ otherwise.

\bigskip\centerline {\bf {B. Dynamics of a single interface}} \smallskip

To understand the evolution of domains in the initially homogeneous
system it will be helpful to determine the dynamics of a single $AB$
interface.  Therefore consider the ``separated'' initial condition
where $n_A(x,t=0)=\overline n_A$ for $x<0$ and $n_A(x,t=0)=0$ for
$x>0$, whereas $n_B(x,t=0)=0$ for $x<0$ and $n_B(x,t=0)=\overline n_B$
for $x>0$ (Fig.~2).  The evolution of such a system provides essential
insights into the nature of the reaction front [11-16], and has proven
useful in understanding spatial structure in \ab\ with homogeneous
initial conditions, since relatively sharp reaction fronts naturally
form as the system evolves.  For the separated system, we will apply a
mean-field approach in which the free motion of the $A$ and $B$
particles is modeled by the inviscid Burgers equation.  Then by
invoking mass balance at the $AB$ interface, we determine the
interface velocity and the spatial distribution of reactants.

Assume that the interface initially at $x=0$ moves from $x$ to $x+dx$
in a time $dt$.  Thus the $B$s within the slab $[x,x+dx]$ will be
completely replaced by $A$s.  Consequently, the incident flux of $A$,
$j_A$, must be sufficient to annihilate all the $B$s which do not
escape from the slab and also repopulate the slab at the final density
of $A$s.  That is,
$$
j_A\,dt=n_A\,dx +R, \eqnai
$$
where $R$ represents the $A$s which will undergo reaction in the slab.
Similarly, the outgoing flux of $B$, $j_B$, includes the initial number of
$B$s in the slab minus those that disappear by reaction:
$$
j_B\,dt=n_B\,dx -R. \eqnbi
$$
These two lead to the mass balance equation,
$$
(j_A (x,t)+j_B (x,t))\, dt = (n_A(x,t)+n_B(x,t))\, dx,
\eqnoi
$$
where $dx=v_{AB}\,dt$, with $v_{AB}$ being the interface velocity.  In
the inviscid Burgers equation these currents are related to the
corresponding particle densities, $n_A$ and $n_B$, by,
$$
j_{A,B} (x,t) =n_{A,B} (x,t) (1-n_{A,B} (x,t)).
\eqnoi
$$

The concentration of $A$ at the interface depends upon whether the
interface itself moves faster or slower than the point $\tilde x$
which defines the beginning of the depletion zone for the $A$
distribution (Fig.~2).  According to the Burgers equation, this
boundary point moves with velocity $\tilde v=1-2 \overline n_A$.  Thus
if the interface moves to the right faster than $\tilde x$, the
hypothesized depletion zone exists and the concentration of $A$ at the
interface coincides with solution of the inviscid Burgers equation.
On the other hand, if $v_{AB}<\tilde v$, the depletion zone does not
appear and the concentration of $A$ is constant throughout the domain.
Thus for the concentration of $A$ at the interface, we have
$$
n_A (x,t)=\cases{(1-x/t)/2 & \quad for ~$v_{AB}\geq \tilde v$ \cr\cr
\overline n_A & \quad otherwise. \cr}
\eqnoi
$$
On the other hand, since the concentration of $B$ remains constant
throughout the domain
$$
n_B(x,t)=\overline n_B.\eqnoi
$$
Substituting Eqs.~(6)-(8) and the relations
$dx/dt=v_{AB}$ and $\tilde v=1-2\overline n_A$ into the
mass balance equation (5) gives,
$$
v_{AB}=\cases{1-2 \overline n_B (\sqrt {2}-1)& \quad for ~
$\overline n_A \geq \overline n_B (\sqrt{2}-1)$ \cr
&\cr
1- {\overline n_A ^2+ \overline n_B ^2 \over
\overline n_A + \overline n_B}& \quad for ~
$\overline n_A <\overline n_B (\sqrt{2}-1).$\cr}
\eqnoi
$$

To check whether these formulae account for the propagation of an $AB$
interface on a microscopic level, we numerically simulated the
\ab\ reaction with  initially separated species which undergo the same
driven diffusive motion.  The agreement between the simulation data
and the model predictions indicate that our phenomenological
description based on mass balance and the inviscid Burgers equation
accurately accounts for the microscopic behavior (Fig.~3).  However,
the actual width of the reaction zone is primarily controlled by
diffusion and therefore its behavior is beyond the scope of an
inviscid model.

\bigskip\centerline {\bf{C. Density decay and domain growth rates}}\smallskip

Let us now return to the original problem of equal initial
concentrations of $A$ and $B$ which are randomly distributed.  As time
increases, a region initially rich in a particular species (due to
initial density fluctuations) will evolve into a single-species
domain.  To determine the time dependence of the size of such a
domain, note that if the domain length is $L$, the initial excess of a
particular species and thus the number of particles remaining within
it after the minority species is annihilated is typically of order
$\sqrt L$.  Consequently, the concentration in a typical domain is of
order [1-5]
$$
n(t) \sim 1/ \sqrt {L(t)}.\eqnoi
$$
On the other hand, from \back1\ the rate of change of the domain
length $dL/dt$ is
equal to the difference of the velocities of its two boundaries,
$$
dL/dt \propto |v_{AB} - v_{BA}|.\eqnai
$$
Also from \back2\ this velocity difference is proportional to the
density difference across the interfaces.  These differences are
typically of order the domain densities themselves.  Thus we conclude
that
$$
dL/dt \sim n_A - n_B \sim n(t).\eqnbi
$$
Combining \backs1\ and \lastn, we predict that the domain size grows as
$L \sim t^{2/3}$, while the density decays as
$$
n(t) \sim t^{-1/3}. \eqnoi
$$

It is also instructive to estimate the leading correction to this
result.  Recall that the difference between the solution of the
inviscid and viscid Burgers equation occurs in the boundary regions
whose widths are proportional to $\sim \sqrt {t}$ (Fig.~1).
Consequently, the inclusion of a viscosity term should lead to a
$t^{1/2}$ correction to the typical domain length, \ie, $L \sim
At^{2/3} + B t^{1/2}$.  Correspondingly, this gives rise to a
correction of order $\sim t^{-1/6}$ to the asymptotic behavior of the
density.  This slowly vanishing correction is a possible reason for
slow convergence of the previous [7] and our simulations to the
predicted asymptotic behavior.

\bigskip\bigskip
\chap{SIMULATION RESULTS}

We performed simulations on a system of $5\times10^5$ lattice sites
for $1.5^{30}\cong 1.9 \times 10^5$ time steps and averaged over 30
initial configurations.  As has been observed previously [7], the
local estimate of the density decay exponent from our data is
systematically decreasing with time (Fig.~4(a)).  An asymptotic value
of $-1/3$ for this exponent is not incompatible with our data.  A
larger scale simulation would help resolve this issue.  Another basic
quantity which also exhibits slow approach to asymptotic behavior is
the average domain length $L(t)$ (Fig.~4(b)).  The local exponent of
this quantity grows systematically with time and appears to approach
the asymptotic value 2/3 that was predicted in our phenomenological
approach.

A deeper understanding of the behavior of the system can be gained by
examining the distribution of particles within a domain.  In the case
of \ab\ with isotropic diffusion, we examine both the
``microcanonical'' and ``canonical'' domain profiles [8].  To
construct the former, we first define the extent of a particular
domain of $A$, say, by the distance between the two $B$ particles
which enclose this A domain.  We then rescale the lengths of
individual domains to a constant value and then superpose the
corresponding rescaled densities.  Finally, it is convenient to plot
the scaled profile $\rho(x/L,t)\,t^{1/3}$, where $\rho(x/L,t)$ is the
local density within a domain at scaled position $x/L$ and time $t$
(see Fig.~5).  These data collapse onto a universal curve, except for
the leading and trailing regions, where the slopes for later time seem
to be steepening.  We suggest that the departure from scaling can be
partially understood by recalling that the boundary layers around each
domain grow as $\sqrt t$, while a domain by itself grows as $t^{2/3}$.
Thus the relative width of the boundary layer decreases as $t^{-1/6}$,
leading to the slope of the scaled density profile as $t^{1/6}$.  To
illustrate this effect, we plot, in Fig.~6, the trailing tail of the
microcanonical domain profile when the ordinate is rescaled according
to $t^{1/6}$.  The apparent existence of scaling according to
$t^{1/6}$ in this subregion suggests that there are really two length
scales for a single domain.  One, growing as $t^{2/3}$, corresponds to
the ``bulk'' domain growth and is described by the inviscid Burgers
equation.  The other, corresponding to the boundary layer, grows as
$\sim\sqrt{t}$ and may be described only by including the higher-order
viscosity term in the Burgers equation.

We have also examined the canonical density profile in which the
densities of all domains are superposed about a common center, but
without rescaling the domains lengths before the superposition.
Strikingly, these profiles are nearly symmetric about a point which is
centered some small distance upstream from the domain center (Fig.~7).
Furthermore, the decay of this profile is quite accurately described
by a simple exponential.

\bigskip\bigskip
\chap{DISCUSSION}

To summarize and to place our results in an appropriate context, it is
helpful to consider the \ab\ reaction when there are arbitrary hopping
anisotropies $\b_A$ and $\b_B$, respectively, for the $A$ and $B$
species.  Thus for a given species, the probability for a particle to
jump to the right is $(1+\b)/2$ and to the left is $(1-\b)/2$.  The
model we have considered thus far corresponds to the extreme case of
$\b_A=\b_B=1$, \ie, particles hop only to the right.  The general case
can be usefully described in terms of the ``phase diagram'' whose axes
are the degree of anisotropy of each species (Fig.~8).

In the continuum limit, it straightforwardly follows that for
arbitrary but equal biases of $A$ and $B$, ($0<\b_A = \b_B
\equiv\overline\b\leq 1$), the Burgers equation, Eq.~(3) will be
modified only by the introduction of the coefficient $\overline\b$
which multiplies the first-order spatial derivative,
$$
\pd {\overline n}t=-\overline\b \,\, \pd{}{x}{[\overline n (1-\overline n)]}
+{{1\over 2} \pnd{\overline n}{x}{2}}.
\eqnoi
$$
Such a coefficient in the equation of motion can be absorbed by a
suitable rescaling of $x$ and $t$, leading to universal behavior in
the hopping anisotropy, as long as this anisotropy is non-zero and the
same for both species.  Hence the density decay exponent along the
line $\b_A=\b_B$ is expected to be equal to $-1/3$, except for
$(\b_A,\b_B)=(0,0)$.  This singular point corresponds to the case of
pure diffusion where the density decay exponent equals $-1/4$,
independent of the existence of a same species hard core repulsion
[1-5,8].

The remainder of the phase diagram corresponds to systems with unequal
biases for the two species.  Applying the arguments of Sec.\ II, the
evolution of the typical domain length is determined by the difference
in the velocities at each interface.  This gives $dL/dt \sim \b_A -
\b_B$, so that the average domain size grows linearly in time.
Consequently the concentration varies as $n(t)\sim 1/\sqrt{L}\sim
1/\sqrt{t}$.  Our simulations of \ab\ with an unequal biases for $A$
and $B$ are consistent with this heuristic prediction.

Thus diffusion-controlled two-species annihilation with same-species
hard-core repulsion has rich variety of kinetic behaviors in one
dimension.  The density decay exponent equals 1/4, 1/3, and 1/2 for
zero, equal, and unequal biases, respectively.  For the first and
third situations, rate equation approaches have failed to give the
correct asymptotic behavior. In contrast, for the equal bias case we
have shown that the Burgers equation approach, supplemented by mass
balance at the interfaces between domains, explains several
characteristics of the system.  More subtle features, such as the
density profile inside the domains, are beyond the simple-minded
approach developed here but might be accessible by a refinement of our
technique.  It is also worth mentioning that macroscopic
characteristics of the system, \eg, the typical domain size $L\sim
t^{2/3}$, naturally arise in other contexts where the Burgers equation
plays the leading role [17].

Finally, it is worth considering the effect of driven motion on \ab\
in greater than one spatial dimension. Numerically, we have
investigated two natural versions: In the NE model, we allow particle
to either move up or to the right with equal probabilities, while in
the NES model, we allow particles to move either up, down, or to the
right with equal probabilities.  For both cases, the density appears
to decay asymptotically as $t^{-1/2}$.  Interestingly, there is a
significant time range where the density decay exponent is clearly
less than $-1/2$.  This same anomaly also appears in \ab\ with
isotropic diffusion of the reactants, however.  This can be ascribed
to the existence of a limited temporal regime where there is some
penetration of particles of one species into a domain of the opposite
species.  This effect is even more strongly pronounced in three
dimensions [18].

Qualitatively, one can estimate the effect of driven diffusion on the
reaction kinetics in arbitrary dimension $d$.  If $L$ is the typical
length of a single-species domain in the drift direction, then its
volume $V$ scales as $Lt^{(d-1)/2}$, since the domain length in the
remaining $d-1$ transverse directions still grows by diffusion.
Repeating now the steps employed in the previous treatment of the
one-dimensional situation, we have $dL/dt \sim n$ and $n\sim V^{-1/2}
\sim L^{-1/2}t^{-(d-1)/4}$.  These can be readily solved to yield
$$
n \sim \cases{t^{-(d+1)/6}& \quad for ~$d\leq 2$ \cr
&\cr
t^{-d/4}& \quad for ~$2<d\leq 4$ \cr
&\cr
t^{-1}& \quad for ~$d>4$.\cr}
\eqnoi
$$
Our simulations are covered by the first line of \last.  When $d>2$,
the longitudinal length predicted by the above argument, $L\sim
t^{(5-d)/6}$, is smaller than the diffusional length scale.  This
indicates that the drift is irrelevant above two dimensions and the
reaction kinetics is the same as two-species annihilation with
isotropic diffusion.

\bigskip\bigskip
\centerline{\bf Acknowledgements}
\medskip
We thank D.~ben-Avraham, S.~Esipov, and V.~Privman for helpful
discussions.  We also gratefully acknowledge ARO grant
\#DAAH04-93-G-0021 for partial support of this research.  As this
paper was being written, we learned of parallel work by Janowsky [19].
Although the spirit of our approaches are compatible, there are some
quantitative disagreements of our respective simulation results.

\vfill\eject

{
\parindent=0.2in
\centerline{\bf References}\smallskip

\refi Ya.~B.~Zel'dovich, A.~A.~Ovchinnikov, \cp 28 215 1978 ;
Ya.~B.~Zel'dovich, A.~S.~Mikhailov, \usp 30 23 1988 .

\refi D. Toussaint, F. Wilczek, \jcp 78 2642 1983 .

\refi K. Kang, S. Redner, \prl 52 955 1984 ; \pra 32 435 1985 .

\refi G. Zumofen, A. Blumen, J. Klafter, \jcp 82 3198 1985

\refi M. Bramson, J. L. Lebowitz, \jsp 62 297 1991 ;
\jsp 65 941 1991

\refi V. Privman, A. M. R. Cadilhe, and M. L. Glasser, preprint.

\refi S.~A.~Janowsky, \pre 51 1858 1995 .

\refi F. Leyvraz and S. Redner, \pra 46 3132 1992 ; see also S.
Redner and F. Leyvraz, in {\it Fractals and Disordered Systems, Vol.
II\/} eds.\ A. Bunde and S. Havlin (Springer-Verlag 1993).

\refi H.~Spohn, {\sl Large-Scale Dynamics of Interacting Particles},
      Texts and Monographs in Physics, Springer-Verlag, 1991.

\refi G.~B.~Whitham, {\sl Linear and Nonlinear Waves},
      Wiley-Interscience, 1974 (Ch.4).

\refi L.~G\'alfi and Z.~R\'acz, \pra 38 3151 1988 .

\refi H.~Larralde, M.~Araujo, S.~Havlin,
      and H.~E.~Stanley, \pra 46 855 1992 ; \pra 46 6121 1992 .

\refi S.~Cornell and M.~Droz, \prl 70 3824 1993 .

\refi E.~Ben-Naim and S.~Redner, \jpa 25 L575 1992 .

\refi B.~Lee and J.~Cardy, \pre 50 3287 1994 .

\refi P. L. Krapivsky, \pre 51 xxx 1995 .

\refi T.~Tatsumi and S.~Kida, \jfm 55 659 1972 ;
      S.~F.~Shandarin and Ya.~B.~Zel'dovich, \rmp 61 185 1989 .

\refi F. Leyvraz, \jpa 25 3205 1992 .

\refi S. A. Janowsky, preprint.
}

\vfill\eject
\centerline{\bf Figure Captions}

\rfigi Asymptotic solution of the Burgers equation for the initial
value problem $n(x,t=0)= A\d (x)$ with (-----) and without ($\cdots$)
the viscosity term.

\rfigi Schematic illustration of evolution of the interface between two
initially separated $A$ and $B$ domains.  The motion of each species
is described by the inviscid Burgers equation.  There is a depletion
zone in the $A$ distribution which begins at $\tilde x$ and extends to
the interface.  In a time $dt$, the interface advances from $x$ to
$x+dx$ and the initially uniform $B$ distribution in the slab
$[x,x+dx]$ is replaced with an slowly varying $A$ distribution whose
density is approximately equal to the $A$ density at the interface.

\rfigi Density profile near the $AB$ interface at
$t=0 (\diamond), t=50 (+), t=100 (\star), t=150 (\times)$, and $t=200
(\bigtriangleup)$, for 300 configurations with $\overline n_A
=\overline n_B = 0.5$.  The calculated positions of the interface are
marked by the $\circ$. The dotted line shows the calculated value of
$n_A=0.207$ for the $A$ density at the interface.

\rfigi Time dependence of the local exponents for: (a) the density
decay exponent, and (b) the domain length exponent.

\rfigi The scaled microcanonical density
profiles at $t=1.5^{18}\cong 1477 (\bigtriangleup), t=1.5^{21}\cong
4987 (\times), t=1.5^{24}\cong 16833 (\star), t=1.5^{27}\cong 56814
(+)$, and $t=1.5^{30}\cong 191750 (\diamond)$.

\rfigi Trailing edge of the microcanonical
density profiles when rescaled by the factor $t^{1/6}$.  The symbols
refer to the same times as Fig.~5.

\rfigi The canonical density profile at $t=1.5^{21}\cong 4987$.

\rfigi ``Phase diagram'' for the density decay exponent for different
hopping anisotropies of the $A$ and $B$ species.

\vfill\eject

\vfill\eject\bye